\documentclass[conference]{IEEEtran}
\IEEEoverridecommandlockouts
\usepackage{adjustbox}
\usepackage{algorithmic}
\usepackage{amsmath, amssymb, amsfonts}
\usepackage{cite}
\usepackage{graphicx}
\usepackage{stackengine}
\usepackage{textcomp}
\usepackage{xcolor}
\usepackage{hyperref}
\usepackage{multirow}
\usepackage{colortbl}
\usepackage [latin1]{inputenc}

\def\BibTeX{{\rm B\kern-.05em{\sc i\kern-.025em b}\kern-.08em
    T\kern-.1667em\lower.7ex\hbox{E}\kern-.125emX}}

\title{Machine Learning for Deferral of Care Prediction}
\author{

\IEEEauthorblockN{
Muhammad Aurangzeb Ahmad$^1$,
Raafia Ahmed$^1$, Dr. Steve Overman$^1$, Patrick Campbell$^1$, \\
Corinne Stroum$^1$, Bipin Karunakaran$^2$
}

\\
1. \textit{KenSci Inc.}\\
2. \textit{Novant Health}\\

\\
\{muhammad, raafia.ahmed, drsteve, patrick.campbell, corinne\} @kensci.com
bkarunakaran@novanthealth.org
}

\begin{document}
\maketitle

\begin{abstract}
Care deferral is the phenomenon where patients defer or are unable to receive  healthcare services, such as seeing doctors, medications or planned surgery. Care  deferral can be the result of patient decisions, service availability, service limitations, or restrictions due to cost. Continual care deferral in populations may lead to a decline in population health and compound health issues leading to higher social and financial costs in the long term. \cite{findling_delayed_2020}. Consequently, identification of patients who may be at risk of deferring care is important towards improving population health and reducing care total costs. Additionally, minority and vulnerable populations are at a greater risk of care deferral due to socioeconomic factors. In this paper, we (a) address the problem of predicting care deferral for well-care visits; (b)  observe that social determinants of health are  relevant  explanatory factors towards predicting care deferral, and (c)  compute how fair the models are with respect to demographics, socioeconomic factors and selected 
comorbidities. Many health systems currently use rules-based techniques to retroactively identify patients who previously deferred care. The objective of this model is to  identify patients  at risk of deferring care and allow the health system to prevent care deferrals through direct outreach or social determinant mediation. 
\end{abstract}
\section{Introduction}
\label{sec:introduction}
Disruption in access to healthcare can have long term negative consequences to a person's health. For example, a study conducted by the BMJ indicated that even a four week treatment delay in certain cancers increases the risk of death by 6$\%$ to 13$\%$ \cite{hanna_mortality_nodate}. Socio-economic factors like physical disability, lack of transportation, limited financial resources etc. have been associated with care deferral, as these factors become barriers to attaining appropriate healthcare \cite{palm_uneven_2021}. A study by Palm et al. found that COVID-19 propagated healthcare deferrals by persons avoiding public transportation and persons without access to their own vehicle \cite{palm_uneven_2021}. At the onset of the pandemic, government and healthcare leaders cancelled many elective procedures and preventative care services, and certain healthcare services  continue to be restricted. Although these measures are deemed necessary, they created a "crisis within the crisis." Healthcare personnel had to urgently implement strategies to mediate this crisis through telehealth access, screening, diagnosis and online treatment consultations \cite{purcell_then_2021}. Even as health systems have re-opened to elective procedures, many patients continued to defer care due to the COVID-19 outbreak \cite{findling_delayed_2020}. Researchers from the National Cancer Institute  estimate ~10,000 excess deaths from breast or colorectal cancer over  ten years due to missed preventative care \cite{sharpless_covid-19_2020}. Care deferral for urgent conditions like stroke and other medical emergencies can also lead to increased in-home deaths \cite{dejong_deferral_2021}. Recent studies have also shown that another impact of the pandemic was a dramatic reduction in admissions of persons with chronic conditions \cite{bhambhvani_hospital_2021} \cite{blecker_hospitalizations_2021}.

Prior to COVID-19, minorities and other vulnerable groups had barriers to accessing healthcare \cite{noauthor_adults_nodate}; the  pandemic has worsened this problem significantly. COVID-19 studies show that care deferral accelerated its spread, especially within vulnerable groups \cite{ponnambalam_understanding_2012}. Care deferral may not be explicitly identified unless patient interviews are conducted to determine the factors driving deferral, such as  cost of care, loss of insurance cover and extent of cover, loss of employment, access to transportation, fear of contracting COVID-19, or perceptions that care  is not needed. Regardless of the deferral reasons, long term effects from condition onset or exacerbation may detrimentally impact a person's health and/or lead to increased healthcare costs. 

In this paper we address the problem of identifying patients at risk for deferring care.  If such patients can be proactively identified then health system user groups, such as  Population Health or Health Equity teams, may conduct outreach and mitigate or address social determinant factors to reduce care deferral. We developed a predictive model that allows the patient population to be risk stratified for deferring well-care encounters, resulting in a health system being able to integrate this predictive model into their workflow. Studies show that limited access to healthcare contributes to ill health and Poverty cycles \cite{fisher_as_2020}. This model aims to generate special attention to minority and vulnerable populations, in order to lessen care disparities and enhance equity. 

Novant Health, with the support of The Duke Endowment, partnered with KenSci to build a Machine Learning system that identifies patients at risk for deferring their well-care encounters. Novant Health  is a healthcare system comprising of 15 hospitals across four states on the East Coast. The Duke Endowment is one of the nation's largest private foundation, investing in communities across North Carolina, while ensuring racial equity. The code for the care deferral model is available in an open source GitHub repository.\footnote{\url{https://github.com/KenSciResearch/CareDeferral}} 

The rest of the paper is organized as follows: In section \ref{sec:related} we describe related work, sections \ref{sec:data} and \ref{sec:exp} we describe the data and the experiments. In addition to
determining the predictive performance, we focus on model explanations and fairness of predictive models in section \ref{sec:fairness}. In healthcare applications, model explanations are an integral part of model usage, as they are used  by healthcare end users to make decisions on individual patient or population action plans.  \cite{ahmad_interpretable_2018}. In section \ref{sec:xai} we summarize observations related to top factors for model explanations where many social determinants of health show up. In section \ref{sec:fairness} we explore a critical component of responsible AI - if and how model fairness varies across cohorts. 

\section{Related Work}
\label{sec:related}
There are several studies which assess the impacts of care deferral on health outcomes, and populations most affected by care deferral. In a 10 year longitudinal study of non-elderly patients with atherosclerotic cardiovascular disease (ASCVD) Khera et al \cite{khera_abstract_2018}, it was reported that one in five patients deferred care due to cost. Jatrana and Crampton \cite{jatrana_financial_2021} observed health deterioration due to care deferral for a longitudinal study in New Zealand and found that cost barriers to doctor visits negatively impacted the mental health of males and young adults. He and Wu \cite{he_towards_2017} observed care deferral due to economic disparities between urban and rural parts of China. Care deferral was prevalent BY migrants to urban areas because they could no longer afford it. Morales et al \cite{morales_bridging_nodate} observed that short term medical missions can greatly reduce care deferral in certain locales.

a growing body of research indicates COVID-19 influenced care deferral for a large segment of the population. De Jong et al \cite{dejong_deferral_2021} highlighted the negative long term hidden effect of COVID-19 on non-COVID conditions due to care deferral. Blecker et al \cite{blecker_hospitalizations_2021-1} observed  substantial changes in hospitalization patterns in New York City in 2020 due to COVID-19 for chronic conditions. Atherly et al \cite{atherly_consumer_2020} observed a marked increase in care deferral associated with COVID-19 and that  telemedicine lead to partially mitigating negative impacts. Reuter et al \cite{reuter_predicting_2022} observed widespread care deferral due to COVID-19 across 28 different European countries.

Powis et al \cite{powis_impact_2021} observed the  negative impact the quality of care and patient outcomes for care deferral among cancer patients, and they predicted that the fallout from cancer care deferral may be observed for years to come. Palm et al \cite{palm_uneven_2021} studied the effects of lack transportation during the early months of COVID-19 and showed that minorities, disabled people, and low-income people were more likely to be impacted. Sun et al \cite{sun_worse_2021} reported worse cardiac arrest outcomes during COVID-19 in Boston could be attributed to patient reluctance to seek care because of fears related to COVID-19. Lastly, a large scale national longitudinal study by Whaley et al \cite{whaley_changes_2020} observed similar patterns of deferred care.

\begin{table*}
\centering
\begin{tabular}{l|l|c}
\hline
\textbf{Feature Category} & \textbf{Example Features} & \textbf{Total Features} \\
\hline
Demographics  &  Age, Ethnicity, Gender  &             9            \\
Diagnosis  &  Diabetes, Heart Disease, COPD  &      41                   \\
Utilization  &  Total ED Encounter, Total Encounters  &         12                \\
Medications  & Anticoagulants Med, Cardiovascular Med &           14              \\
Labs  & Hemoglobin A1C, Sodium, BUN   &   3 (22 aggregated labs)   \\
SDOH  & $\%$ Unemployed, $\%$ No Vehicle, $\%$ Minority  & 21 \\                       
\hline
\end{tabular}
\caption{Feature sets used for prediction}
\label{tab:features}
\end{table*}

\section{Data \& Methodology}
\label{sec:data}
Novant Health's Electronic Health Record (EHR) was used to develop, experiment with and validate a predictive model for Care Deferral. An additional source of data that augmented the EHR data was the Social Vulnerability Index (SVI) developed by CDC and ATSDR \cite{flanagan_social_2011}. Our inclusion criteria consisted of patients who had well-care visit during a prior 2-year period.  We define well-care Visits according to a value set in the Healthcare Effectiveness Data and Information Set (HEDIS)\footnote{The measure results have not been certified via NCQA's Measure Certification Program and constitute "\textbf{Adjusted, Uncertified, Unaudited HEDIS Rates}". NCQA disclaims all use or interpretation of the results. The Healthcare Effectiveness Data and Information Set (HEDIS) is a registered trademark of NCQA. The HEDIS measure for Adult Access to Preventative/Ambulatory Health Services (AAP) measure reports health plan member rates of ambulatory or preventative care visits} \cite{noauthor_adults_nodate}, and our use of this value set ensures alignment with other healthcare analyses. This value set includes a set of CPT, HCPCS and Diagnosis Codes for annual physical exams, telemedicine visits, office visit, Medicare wellness visits and similar encounters. Care visits are separate from other preventive services such as standard vaccinations or screenings such as Mammograms or Colonoscopies. In other applications of this model, the value sets predicted could be swapped or added to learn about other types of deferred preventive care.  
Therefore, we thus only selected patient cohorts who had at least one well-care encounter. Consequently, patients who never received well-care (e.g., patients who had some encounter in the EHR but no well-care encounter) were excluded from this cohort. We determined that simple rules (e.g., based on utilization) could find those patients who have never had a well-care visit in the system.

During the selection process for our inclusion criteria, we identified five potential cohorts / sub-cohorts the model could learn from. 
\begin{itemize}
    \item All community members in Novant Health's Service Areas    
     \item Members from the Novant Health service areas with no encounter at Novant Health (these persons do not show in the EHR)
     \item Members from the Novant Health service areas with at least encounter at Novant Health (these patients do show in the EHR)
     \item Patients with at least one well-care encounter
     \item Patients with no well-care encounter
\end{itemize}

\section{Experiments \& Results}
\label{sec:exp}
\subsection{Problem Setup}

\begin{figure*}
    \centering
  \includegraphics[width=6.5in]{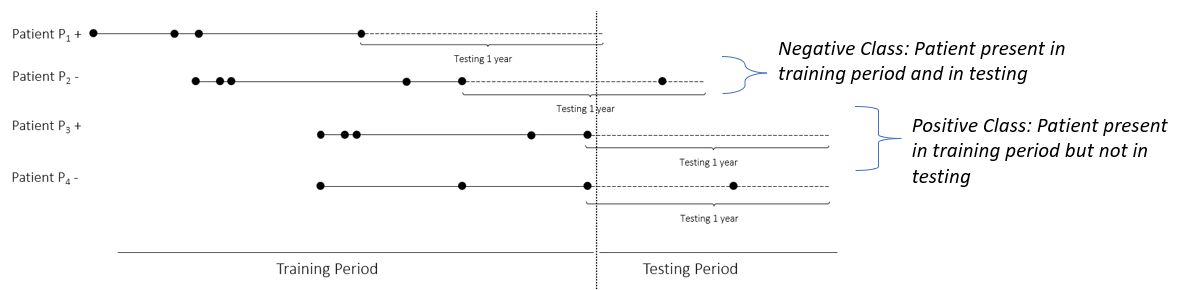}
  \caption{Definition of Positive and Negative examples for Care Deferral}
  \label{fig:def}
\end{figure*}

We posit the problem of predicting care deferral as a binary classification problem. A positive instance is defined as a patient who has had a well-care encounter in the training period but \textit{not} in the test period. A negative instance is when the patient has had a well-care encounter in both the training and the test period. We note that an instance is considered to be a negative instance if there is even a single well-care follow up encounter within the span of the test period as long as it is within a year of last well-care visit in the training period. The training period spans from January 2017 to December 2018 while the test period spans from January 2019 to December 2019. The data from 2020 was not employed since the time frame was non-representative of normal patterns of care deferral due to the pandemic. The problem setup is visualized in Figure \ref{fig:def}. 

After applying the criteria, the final cohort consisted of 779,506 patients of which 322,667 (41.39$\%$) deferred care and 456,839 (58.61 $\%$) did not defer care. Representation across gender is fairly balanced.  The patient data that we used comes from EHR encounter data, which includes demographics, diagnoses, utilization,  medications, and labs. Publicly available data for Social Vulnerability Index, as described in the previous section, was also used. We note that there is some loss of granularity since all EHR derived variables are characteristics of the individual while the features that are derived from SVI are linked to the locale (county) that the patient resides in. We acknowledge that the Census Tract can also be utilized, as it is provided in the SVI, however we did not have the patient census tracts available during the timeframe of this project. Social Determinants of Health Features utilizing census tracts is preferred, as counties can be quite large and may not represent the patient well.

\subsection{Model Performance}
In building the model, we compared a standard set of standard models: Logistic Regression, Ada Boost, Random Forest, Naive Bayes, SVM, XGBoost, and Decision Trees. A summary of prediction results is given in Table \ref{tab:results} for the best model, XGBoost, and the baseline models. From the results it is clear that the best set of results are obtained from the XGBoost model with auto-tuning of hyperparameters. We also use a statistical baseline where the prediction is done at random with respect to the relative distribution of the two classes. 

\begin{table}[]
\centering
\begin{tabular}{l|c|c}
\hline
\textbf{Metrics} & \textbf{XGBoost} & \textbf{Baseline (Statistical)} \\
\hline
Accuracy & 0.52 & 0.73\\
Specificity & 0.59 & 0.74\\
Precision & 0.41 & 0.72\\
Recall & 0.41 & 0.65\\
F-Score & 0.41 & 0.68\\
AUC & 0.50 & 0.79\\
\hline
\end{tabular}
\caption{Model Performance}
\label{tab:results}
\end{table}

\subsection{Model Explanation}
\label{sec:xai}
The goal of predicting deferral of care is to employ the model as a tool to reduce the number of such instances. Thus, it is important to understand what factors may be driving patients to defer care. One way to accomplish this is via explanation of model predictions. While model explanations are not causal in nature, a domain expert can still look at the model explanations and determine what factors may be driving a patient to defer care. These insights could then be used to create policies that can help reduce instances of deferral of care. We employed the SHAP framework \cite{lundberg_unified_2017} for model explanations. A condensed list of top 15 risk factors and their average values of those risk factors for the cohort are given in Figure \ref{fig:explain}.

\subsubsection{Utilization Features}
\begin{figure}
 \centering
  \includegraphics[width=3.5in]{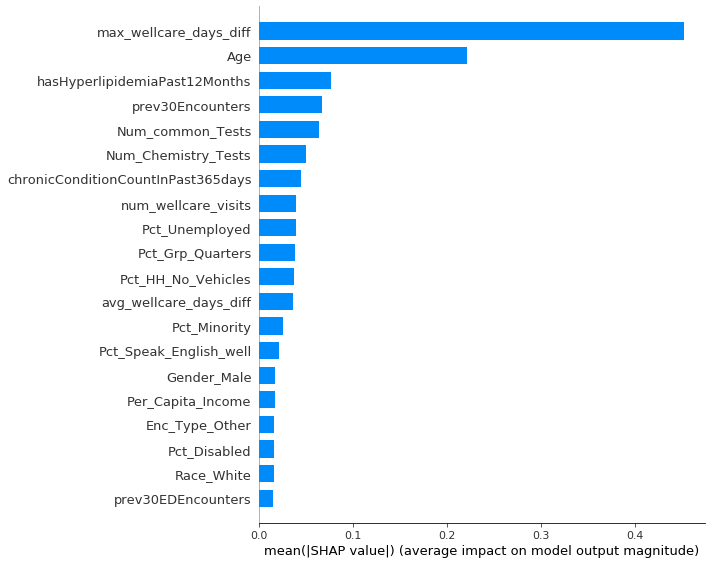}
  \caption{SHAP values for Care Deferral Model}
  \label{fig:explain}
\end{figure}

From the model explanations in \ref{fig:explain}, a number of features stand out: maximum difference between previous well-care encounters appear to be an important factor. We see high values for Max days between well-care encounters impacts patients deferring care. Other features associated with deferring well-care were the average days difference between well-care encounters and  number of prior well-care encounters. This illustrates that past diligence in keeping up with well-care visits is a good indicator of future compliance for well-care visits (e.g., High number of well-care visits, indicates patient is less likely to defer care). Additional utilization factors that associate with deferring well-care are overall number of encounters in the previous 30, number of days from the last well-care encounter, and the total number of ED encounters. 

\subsubsection{Lab Features}
The feature set consisted of two labs features (chemistry labs and common labs), both of which show up in the most associated feature lists. An earlier iteration of the model had individuals labs encoded as separate features, however, combining them led to slight improvement in model performance. Both labs features are binary functions i.e., the value is one if the test was done and zero otherwise. The common labs feature consists of the following labs: Liver function tests, Renal Panel, Lipid labs, Thyroid function tests, urine tests and blood tests. The chemistry labs consist of the following: Albumin, Sodium, Potassium, Bilirubin, Blood urea nitrogen (Bun), and Alkaline Phosphate.

\subsubsection{Social Determinants of Health Features}
 A large number of social determinants of health (SDoH) show up in the list of most important features. Most of these features are aggregate features at the county level, which are linked to the individual patients residing in that county. The SDoH features that are in the top list of features are:  Percentage of households with no vehicles, percentage of people unemployed, percentage of people living in group quarters, percentage of people who are minorities, percentage of disabled people, percentage of people living in poverty, per capita income, and percentage of people who speak English less than well.

\begin{table*}[h!]
\centering
\begin{tabular}{c|c|c|c|c|c|c|c|c|c}
\hline
\multicolumn{1}{l|}{\textbf{Metric}} & \textbf{Measure} & \textbf{Female} & \textbf{Asthma} & \textbf{COPD} & \textbf{HF} & \textbf{CVD} & \textbf{PUD} & \textbf{PVD} & \textbf{MLD} \\
\hline
\multirow{9}{*}{Group Fairness} & AUC Difference &  0.0295 & 0.0181 & 0.0188 & 0.0292 & 	0.0361 & 	0.0206   & 0.0315	 & 0.0283 \\
 & Balanced Accuracy Difference  & 0.0206 & 0.0279 & 0.0296 & 0.0362 & 0.0518 & 0.0248 & 0.0483 & 0.0371\\
 & Balanced Accuracy Ratio & 1.0315 & 1.0438 & 1.0465 & 1.0576 & 1.0845 & 1.0388   & 1.0784	 & 1.0592 \\
 & Disparate Impact Ratio & 0.9457 & \cellcolor{yellow}1.3274 & \cellcolor{yellow}1.3792 & 	1.1972 & \cellcolor{yellow}1.6324	 & \cellcolor{yellow}1.4083 &  \cellcolor{yellow}1.6933 &  \cellcolor{yellow}1.3443 \\
 & Equal Odds Difference  & -0.0439 & \cellcolor{yellow}0.1265 & \cellcolor{yellow}0.1413 & \cellcolor{yellow}0.1165   & \cellcolor{yellow}0.2117 & \cellcolor{yellow}0.1351 &  \cellcolor{yellow}0.2131  &   \cellcolor{yellow}0.1348 \\
 & Equal Odds Ratio & 0.8481 & \cellcolor{yellow}1.3351 & \cellcolor{yellow}1.4034 & \cellcolor{yellow}1.2407 & \cellcolor{yellow}1.6405 & \cellcolor{yellow}1.4582 &  \cellcolor{yellow}1.7274 &  \cellcolor{yellow} 1.2885  \\
 & Positive Predictive Parity Difference & 0.0436 & 0.0226 & 0.0197 & -0.0179 & 0.0267 & 0.0182   & 0.0308 & 0.0543\\
 & Positive Predictive Parity Ratio & 1.0702 & 1.0366 & 1.0317 & 0.9727 & 1.0436 & 1.0293   & 1.0506   & 1.0928\\
 & Statistical Parity Difference & -0.0231 & \cellcolor{yellow}0.1062 & \cellcolor{yellow}0.1197 & 0.0689   & \cellcolor{yellow}0.1643 & \cellcolor{yellow}0.1211   &   \cellcolor{yellow}0.1734  &  \cellcolor{yellow}0.1077\\
\hline
\multicolumn{1}{l|}{Individual Fairness}  & Between-Group Gen. Entropy Error & 0.0001 & 0.0003   & 0.0004 & 0.0001   & 0.0003 & 0   & 0.0003   & 0.0001\\
\hline
\multicolumn{1}{l|}{Data Metrics}  & Prevalence of Privileged Class ($\%$) & 60 & 13 & 16  & 2   & 2 & 1 & 4   & 4   \\
\end{tabular}
\label{tab:fair}
\caption{Measurement of Model Fairness Across Cohorts}
\end{table*}

\subsection{Model Fairness}
\label{sec:fairness}
There are multiple notions of fairness of machine learning models that map to various measures and metrics of fairness \cite{ahmad_fairness_2020}. We computed fairness of our model across demographics, socioeconomic factors as well as comorbidities. The focus of fairness analysis is to determine differences across cohorts for protected attributes like race, gender, ethnicity, socioeconomic status etc. In addition to measuring differences across protected attributes, we also measured differences in model fairness across comorbidities e.g., Diabetes, COPD etc. This analysis was done to highlight that model performance across conditions can vary and predictive models may adversely affect patients who may have certain comorbidities.

The results are summarized in Table \ref{tab:fair}. Due to limitations in space we only show a subset of variables for which differences across cohorts were observed. The metrics given in the tables either compute ratios or statistical differences between the predictive performance across cohorts of interest. In general, a statistical ratio measure $\theta$ is considered relatively fair if $0.8 \leq \theta \leq 1.2$. And any fairness measure of statistical difference $\phi$ is considered fair if $-0.1 \leq \phi \leq 0.1$. No noticeable differences were found in fairness metrics across different demographic groups or across socio-economic status. We however note that there are differences in model performance across co-morbid condition groups of Asthma, COPD (Chronic Obstructive Pulmonary Disease), HF (Heart Failure), CVD (Cardiovascular Disease), PUD (Peptic Ulcer Disease), PVD (Peripheral Vascular Disease), and MLD (Mild Liver Disease). Differences in model fairness been extensively documented in literature when evaluated across chronic conditions and vulnerable populations.  \cite{ahmad_fairness_2020}.

\section{Discussion}
\label{sec:discuss}
Many health systems currently use rules-based techniques to retroactively identify patients who have deferred care. The objective of this project is to \textit{proactively} predict which patients are at risk for not receiving preventative care services. Novant Health, along with many other healthcare systems, is concerned about care deferral, especially among historically marginalized populations. The risk scores from this model are to be used to prioritize patient lists for individual outreach and cohort mitigation strategies. These predictions are used with SDoH data to identify community factors that impact specific populations, such as public transportation. The model explanations indicate that social factors play an important role in deferral of care predictions. This observation also validates the hypothesis that care deferral is driven by socio-economic factors for a segment of the population. 

To optimally reduce well-care deferral, care managers, healthcare provider teams, and population health professionals will need to work together to not only identify and reach out to high-risk patients, but also to implement strategies to encourage patients to follow-up on their preventative care visits, expand remote options, and to connect patients to appropriate community resources where necessary. There may be additional constraints when applying predictive models in practice e.g., limitations in available resources may imply that precision may need to be maximized at the cost of recall. On the other hand if the goal is maximizing the reduction of deferral of care then the opposite trade-off i.e., maximize recall even if precision suffers is desirable. Both scenarios can be satisfied by adjusting the threshold for prediction for any given classifier.

\section{Conclusion}
\label{sec:conclusion}
In this paper we considered the problem of predicting care deferral  for well-care visits, especially among vulnerable populations. While care deferral is not a new crisis,  COVID-19 has exacerbated care deferrals as many people were forced to postpone elective care or forgo health screenings. \cite{dejong_deferral_2021}. Our results indicate that machine learning can be used to risk-stratify patients in danger of deferring care and a subset of care deferral is associated with socioeconomic factors. In the future we would like to extend the analysis of care deferral during the COVID-19 pandemic and how it differs from baseline rates of care deferral as it may take some time for healthcare usage patterns to return to normal, if at all \cite{powis_impact_2021}. We would also like to extend this analysis to other value sets to learn about care deferral for vaccinations and preventative cancer screenings.

For future work we recommend focusing on feature refinement e.g., usage of more granular SDoH features, ideally at the patient level. For cases where such data may not be available to use Census Tract level data to understand the neighborhoods that the patients reside in. Additional features that may be helpful in care deferral prediction are  (a) mental health related, (b) medications, (c) Compliance, and insurance. Lastly, well-care is one value set  to consider for this project. Other preventative care procedures can be implemented in this model, such as cancer screenings or vaccinations. 
A study of the change during the following three time periods: before, during and after COVID-19 may be valuable in understanding baseline utilization of preventive vs ambulatory care services to patients to help in better resource allocation and preventing care deferral.

\section{Acknowledgment}
The project was funded by a grant from The Duke Endowment. The project was done as partnership between Novant Health and KenSci with the support of The Duke Endowment.
\bibliographystyle{IEEEtran}
\bibliography{references}
\end{document}